# Thermal electric effects and heat generation in polypyrrole coated PET fabrics.

J. Avloni[(1)], L. Florio[(2)], A.R. Henn[(3)] and A. Sparavigna[(4)]

[(1)] Eeonyx Corporation, 750 Belmont Way, Pinole, CA 94564, USA
[(2)] Laboratorio Tessili Innovativi, V. Rosselli 2 , 13900 Biella, Italy
[(3)] Marktek Inc.,13621 Riverway Dr., Suite H, Chesterfield, MO 63017, USA
[(4)] Physics Department, Politecnico di Torino,C.so Duca degli Abruzzi 24, 10129 Torino, Italy

* Author for correspondence: E-mail: amelia.sparavigna@polito.it

**Abstract**
Polypyrrole chemically synthesized on PET gives rise to textiles with a high electric conductivity, suitable for several applications from antistatics to electromagnetic interference shielding devices. Here, we discuss investigations on thermal electric performances of the polypyrrole coated PET in a wide range of temperatures above room temperature. The Seebeck coefficient turns out to be comparable with that of metal thermocouple materials. Since polypyrrole shows extremely low thermal diffusivities regardless of the electrical conductivity, the low thermal conductivity gives significant advantage to the thermoelectric figure-of-merit *ZT*, comparable with that of some traditional inorganic thermoelectric materials. The heat generation is also investigated for possible heating textile devices.

## Introduction

Till recently, the route to mimic metallic conductivity was either to insert into a inherently insulating resin a conductive filler or to coat a plastic substrate with a conductive metal solution [1]. In this way, conductive fibers can be prepared to obtain conductive fabrics, or fabrics already produced, can be metallized with conductive coatings. Metal coated textiles are very interesting materials because they generally show a high electromagnetic interference shielding effectiveness EMI-SE [2,3].

Actually, rather innovative textile materials can be obtained when produced with intrinsically conducting polymers (ICPs). These textiles are able to absorb as well as reflect electromagnetic waves, and then can exhibit certain advantages over metallic materials. ICPs are conjugated polymers, with alternating single and double bonds in the polymer backbone, a necessary condition for charge carriers to move freely along the chain when doping is provided. The most prominent ICPs in EMI-SE are polypyrrole and polyaniline, where electrical conductivity can have values comparable to those observed for poorly conducting metals and alloy [2].

In the mid-1970s, the first polymer capable of conducting electricity was discovered in a new form of polyacetilene. The announcement of this discovery quickly reverberated around scientific community, and the intensity of the search for others magnified dramatically [4,5,6]. Among the first commercial products incorporating conductive polymers there was Contex®, a line of conductive textile products originally manufactured by Milliken [7], starting around 1990, and now produced by Eeonyx Corp., under the tradename of EeonTex ™. Polyester fibers are coated with a conjugate conductive polymer, the polypyrrole, and woven to create an antistatic fabric.
Here, we will shown the results of measurements of thermal electric properties of a polypyrrole coated fabrics, in particular the Seebeck coefficient, which turns out to be comparable with that of metal thermocouple materials. Because polypyrrole shows extremely low thermal diffusivity, a significant advantage in the thermoelectric figure-of-merit is obtained. The value of the figure-of-merit is comparable with that of some traditional inorganic thermoelectric materials [8]. We will also discuss the heat generation of the fabrics for applications in textile heating systems.

## Samples

Polypyrrole is one of the intrinsic conducting polymers very promising for wide thermoelectric applications because of it several attractive properties, such as easy preparation with low costs. Actually, considerable attention is paid to obtain thermoelectric effects from advanced coating technologies, for low weight but high reliability systems. The thermoelectric cooling is considered as a good strategy in the semiconductor electronic chips operating at very high frequencies, where the thermal management becomes crucial.
A proper polypyrrole coated textile device could represent a possible solution for heating and cooling and for temperature monitoring. As previously told, one of the first commercial textile products incorporating conductive polypyrrole (PPy) was the Contex conductive textile. More recently, Contex-like textiles, with a modified PPy coating have been commercially developed that are more conductive and thermally stable. While imparting electrical conductivity and a dark colour to the substrates, the coating process barely affects the strength, drape, flexibility, and porosity of the starting substrates.
For the measurements discussed in this paper, we use a polypyrrole coated PET fabric which was prepared similarly as described previously in [9,10], with raw chemicals purchased from Sigma-Aldrich and used without further purification. Stochiometric molar ratio of organic acid dopant, anthraquinone-2-sulfonic acid to pyrrole-monomer (i.e., 0.33:1) was used to ensure complete doping level. The molar ratio of polymerisation catalyst, iron(III) nitrate, to monomer (pyrrole) equal to 2.3 mol/mol was used for all reactions. The macroscopic texture of this Contex-like fabric is shown in Fig. 1, a net useful for application in heating systems, as we shall demonstrate in following discussion. In Fig.1, a commercial metallized fabric (Leno Ni/Ag coated Nylon) is shown for comparison. Simultaneous *in-situ* polymerisation and deposition of conductive polypyrrole leads to production of conductive, smooth and uniform coating with thickness under 1 $\mu m$, according to transmission electronic microscope measurements (see Fig.2 and 3). The electrical DC surface resistance was measured by using a four-in-line point probe in combination with computerised Loresta-AP meter from Mitsubishi Petrochemical Co., LTD. Surface resistivity at room temperature of conductive fabrics. The PPy/PET fabric has a DC surface resistivity of 306.0 $\Omega / sq$ and the Leno of 0.22 $\Omega / sq$.

As observed in Ref.11, it is possible a formation of insoluble polymers in the bulk solution and on the surface of the substrate simultaneously. The bulk polymerization produces dendritic polymer particles in the solution and the surface polymerization forms a polymer film on the substrate surface. Some of the bulk polymer precipitates on the surface of the substrate and then the SEM analysis shows these particles on the fibers.

**Thermal electric effects.**

Few researches on organic materials for thermoelectric applications have been reported, probably because of their unattractive electronic transporting characters. But recently electrically conducting organic polymers with aromatic structures have attracted a great attention because of electrical properties and considerable thermal stability. Among them, polyaniline and polypyrrole are well studied due to feasible background for wide application to electronic devices and sensors [12,13,14]. The experimental results of systematic investigations on thermoelectric performances of polyaniline and polypyrrole films in a wide temperature range above room temperature have been reported.

Supposing for polymeric compounds a low thermal conductivity, we can obtain significant advantage of the thermoelectric figure-of-merit. Let us remember that figure-of-merit is defined as $ZT = (S^2 \sigma / \kappa) T$, where $S$, $\sigma$, $\kappa$ and $T$ are Seebeck coefficient, electric conductivity, thermal conductivity, and absolute temperature respectively. The value of $ZT$ for polypyrrole is comparable with that of some traditional inorganic thermoelectric materials, for instance $FeSi_2$ [7]: in polypyrrole films, the maximum value is $10^{-2}$ at $423 K$ [15].

We investigated the behaviour of resistivity as a function of temperature. Starting from room temperature, the resistance of a PPy/PET sample placed in a thermostage, was checked till a temperature of 70°C. The resistance behaviour with temperature is typical of a semiconductor with the resistance decreasing linearly as the temperature increases. At room temperature, the resistance was of 172 $\Omega$ in a sample with a length of $4 \, cm$, composed by 10 yarns, each yarn with a diameter of 0.05 cm. At 70°C, the resistance of the sample was of 145 $\Omega$. Assuming a thickness in polypyrrole coating of around 1 $\mu m$, the electric conductivity σ turns out to be $10^4 \, \Omega^{-1} m^{-1}$. This estimation is in good agreement with the value of $1.7 \times 10^4 \, \Omega^{-1} m^{-1}$, given in Ref.16.

Thermoelectric Seebeck coefficient ($S$) and its temperature dependence were determined by connecting a stripe of PPy/PET fabric $0.5 \, cm$ wide with a copper wire. The two materials are electrically connected by the pressure of a very small silver clip, insulated from the junction. The hot junction was placed in the thermostage with a reference Chromel/Alumel thermocouple and the cold junction between the PPy/PET stripe and copper was thermally anchored at room temperature (26 °C). The same anchoring was used for the cold junction of the Chromel-Alumel thermocouple (a diagram of the experimental set-up in Fig.4). Variations in monitored room temperature during measurements were negligible (around 5%).

In Figure 5, the behaviour of the electro-motive force measured for two such PPy/Copper thermocouples is given as a function of temperature difference $\Delta T$ between the actual hot junction temperature and the room temperature. Assuming a value of Copper e.m.f. vs. Platinum of 0.0076 $mV/°C$ [17], we can estimate a value of 0.0133 $mV/°C$ for the PPy vs. Platinum e.m.f. and positive.

To obtain the figure of merit ($ZT$), we estimate the PPy thermal conductivity in the following manner. Thermal diffusivity of polypyrrole films was measured by a laser flash method in Ref. [18]. Polypyrrole films exhibit extremely low thermal diffusivity of $1.3 \times 10^{-2}$ $cm^2 s^{-1}$ at room temperature. The very low thermal diffusivity of polypyrrole films is originated in the lattice structure, in particular from a dominant amorphous character of the chain structure. An amorphous structure strongly reduces the thermal phonon transport, because strong phonon scattering mechanisms appear [19,20]. Specific heat capacity of the polypyrrole film is $0.4\ J\ g^{-1} K^{-1}$, which is higher than those of inorganic materials but still in the range of those of organic polymers, with a positive temperature dependence [18]. The thermal conductivity of polypyrrole film turns out to be $0.2\ Wm^{-1} K^{-1}$. This low thermal conductivity is at least in one order of magnitude lower than that of the best inorganic thermoelectric materials. With the estimates $\sigma = 10^4\ \Omega^{-1} m^{-1}$, $\kappa = 0.2\ Wm^{-1} K^{-1}$, $T = 300\ K$, our measurements on PPy give $ZT = 3. \times 10^{-3}$ in agreement with data on polypyrrole films [15].

The Seebeck electromotive force of Leno Ni/Ag coated Nylon connected with Cu was also measured and the behaviour is shown in Fig. 5 (curve b). The same figure shows that a thermocouple built with Ni/Ag/Nylon and PPy/PET can give the higher electromotive force (curve a). We have also prepared a thermocouple with PPy/PET and a yarn composed of commercial carbon fibers (curve c).

As shown by the measurements here reported, PPy coating can be successfully used with other conductive yarns, for instance Copper, Ni/Ag coated yarns or carbon fibers, obtaining stable textile thermopiles, providing significant thermopowers. Seebeck effect of PPy with Aluminum is repoterd in Ref.[21]. Contex, or other polypyrrole coated textiles, can then represent a valid solution to the problem of exhausted heat, generated from engines and in power plants, to avoid environment thermal pollution. A challenging strategy for textile industry can then be development of new textiles for a partial recover of heat in electric energy.

**Heat generation.**

An adjustable Variac power supply was used to generate an AC current/voltage over the fabric. Voltage and current were monitored by Keithley voltmeter and amperometer. A square shape fabric ($6\ cm \times 6\ cm$) was positioned between two pressed electric contacts (a diagram of the arrangement in Fig.7). The temperature rise was measured using an Omega infrared thermometer, placed to control the center of the sample. In Fig.8, the behaviour of the temperature as a function of the current is given, with the rise of the voltage. According to the power law, the maximum theoretical power achieved from the fabrics is: $P = V I$, where $P$ is the power developed and $V, I$ the voltage and current. The AC current frequency is 50 Hz. In Fig.9, the power and the impedance as a function of current are shown.

In Ref.10, the power density per unit area is assumed to be: $P = V^2 / R_S l^2$, where $R_S$ is the surface resistance and $l$ the size of the sample. Our highest value is 370 $W/m^2$ in agreement with the value obtained in Ref.10.

As the studies demonstrate, a Seebeck effects can be achieved by using a PPy conducting coating a PET fabric. According to the Kelvin relation between the Seebeck $S$ and the Peltier $\Pi$ coefficients, $\Pi = S T$, we can also imagine application in cooling devices of polypyrrole coated fabrics. In fact, the effect is strongly dependent on the design of the cooling cell and in the number of junctions utilized in it. With the PPy coated PET fabric, it is possible to easily make heating fabrics. Since the coating with polypyrrole is possible on many different fibers [22], the potential applications of polypyrrole in the building of heating pads is relevant. We suggest that the PPy-coated fabric used in this study may be practically useful for many applications, including flexible, portable surface-heating elements for medical or other applications.

**Aknowledgement**
The authors are indebted with Angelica Chiodoni for the SEM analysis.

22. D.T. Seshadri and N.V. Bhat, Synthesis and properties of cotton fabrics modified with polypyrrole, Sen'I Gakkaishi 61(4), 103 (2005)

**Figure captions**

Fig.1 Polypyrrole coated PET net and the Leno Ni/Ag coated nylon fabric. The image sizes are $4.3\,cm \times 5\,cm$.

Fig.2 Scanning electron microscope SEM images of the samples with polypyrrole coated fibers. The dust is due to polymer particles in solution, deposited on the net.

Fig.3 SEM images of PPy coated fibers.

Fig.4 The diagram of the experimental set–up for measuring the Seebeck effect. The electromotive force of two thermocouples are compared: a thermocouple is used as a reference to determine the temperature in the thermostage, the other to determine the unknown electromotive force of the material.

Fig. 5 Electromotive force measured for a PPy-PET/Copper thermocouple as a function of the temperature difference $\Delta T$ between the actual hot junction temperature and the room temperature.

Fig. 6 Electromotive force measured for Ni/Ag/Nylon/Contex (a), for Ni/Ag/Nylon/Copper (b), and for PPy/PET/Carbon Fibers as a function of the temperature difference $\Delta T$ between the actual hot junction temperature and cold junction at room temperature.

Fig.7 The diagram of the experimental set-up for measuring the heating effect of a textile. Current and voltage across the sample must be monitored.

Fig.8 Behaviour of the voltage and temperature of the PPy coated sample as a function of the current, measured with voltmeter and thermometer.

Fig.9 Behaviour of the impedance and power developed by the PPy coated sample as a function of the current. The values of impedance and power are estimated from data of Fig.8.

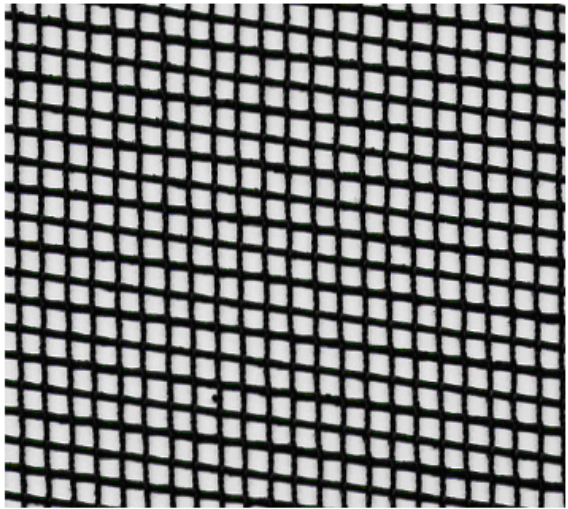
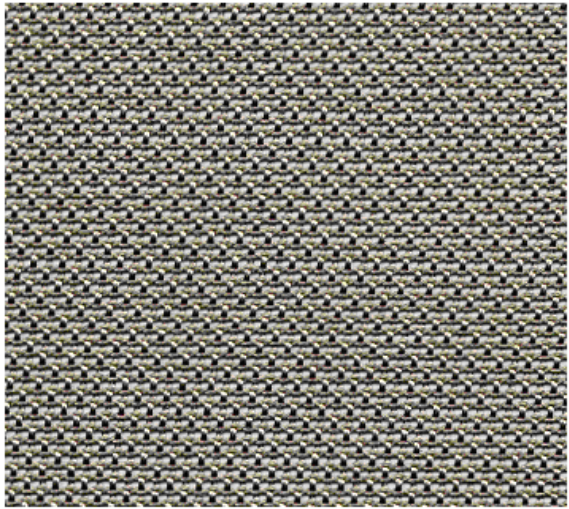

Polypyrrole coated PET

Ni/Ag coated Nylon

FIG.1

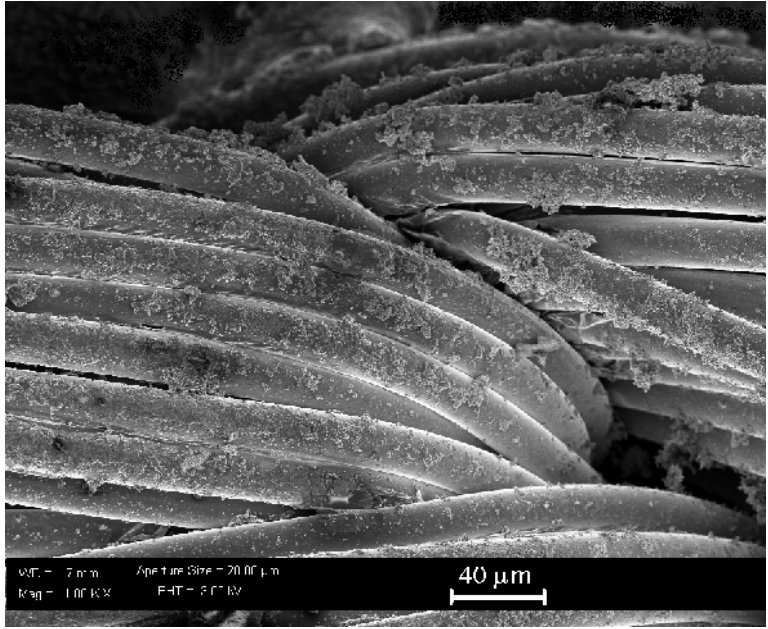

Fig.2

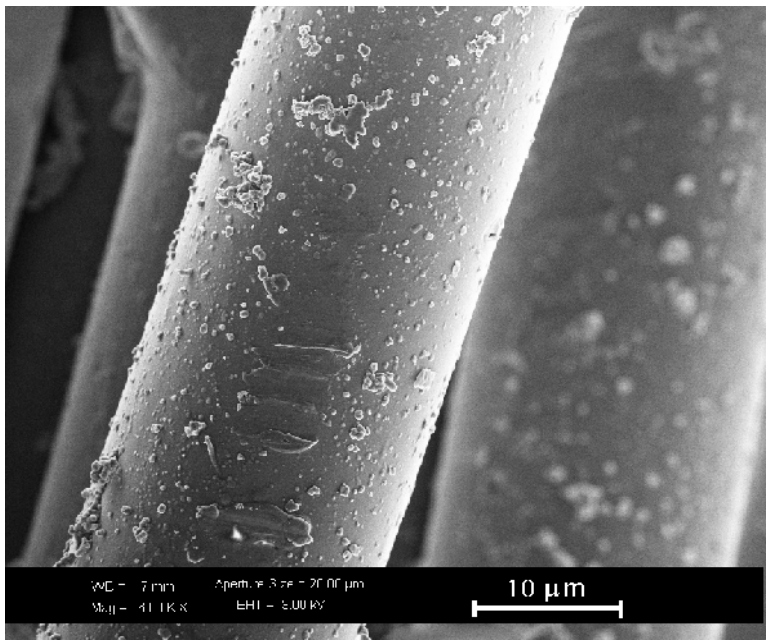

Fig.3

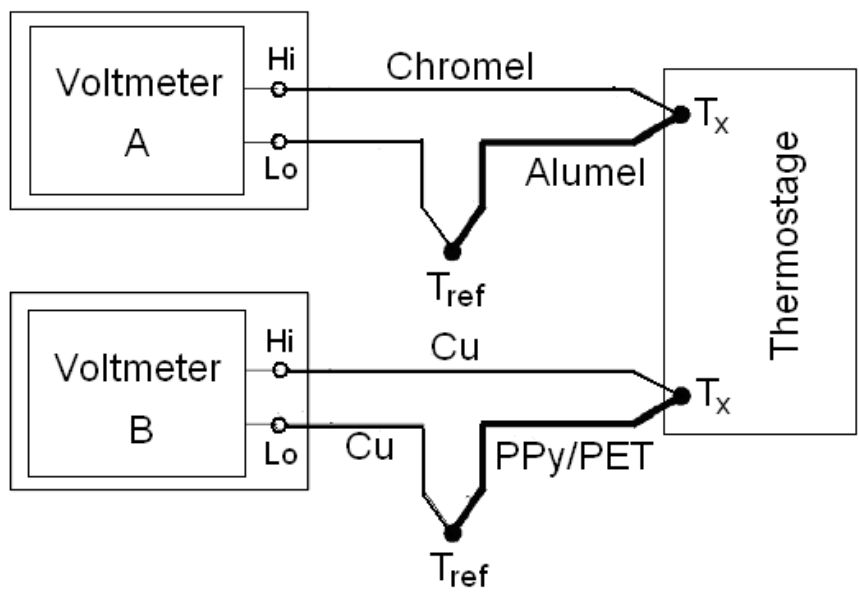

Fig.4

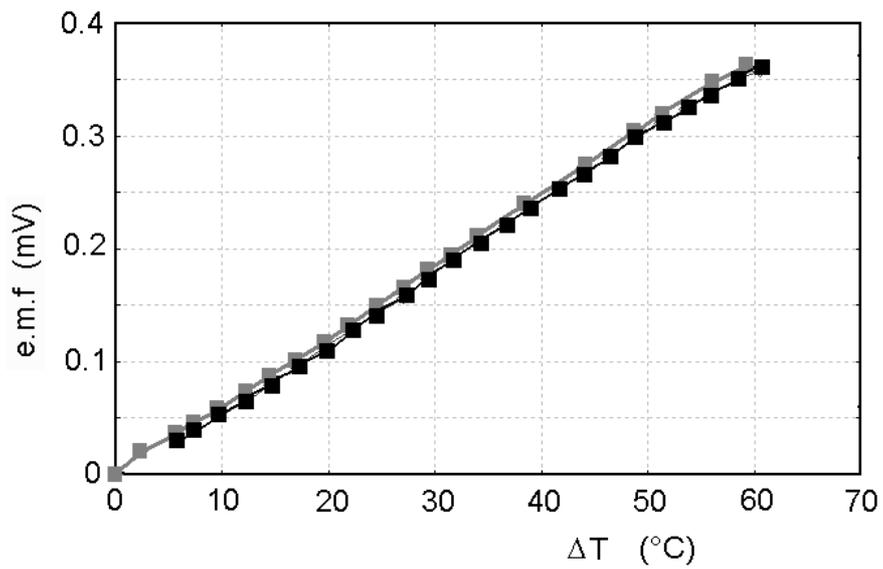

Fig.5

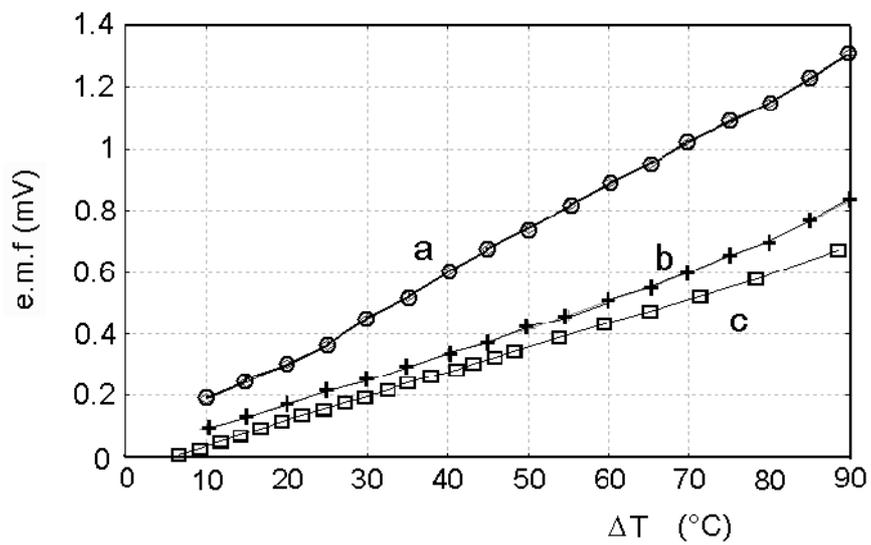

Fig.6

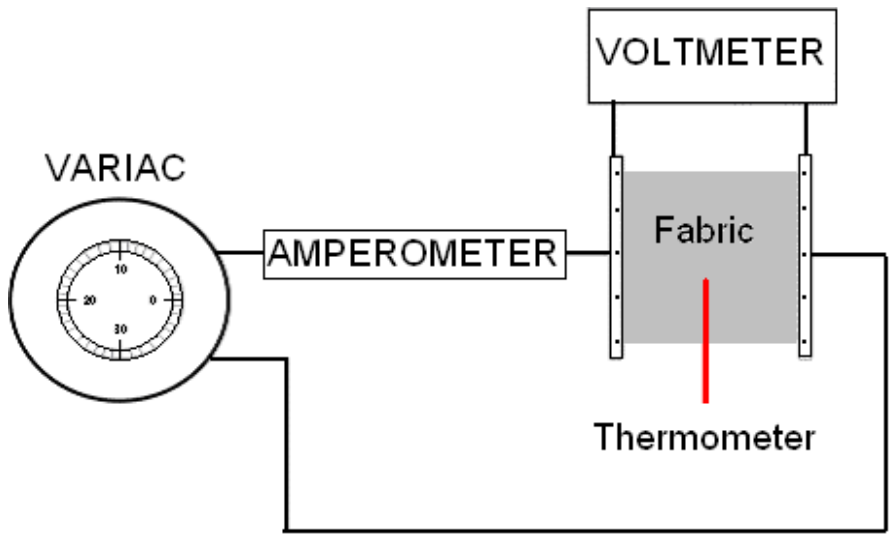

Fig.7

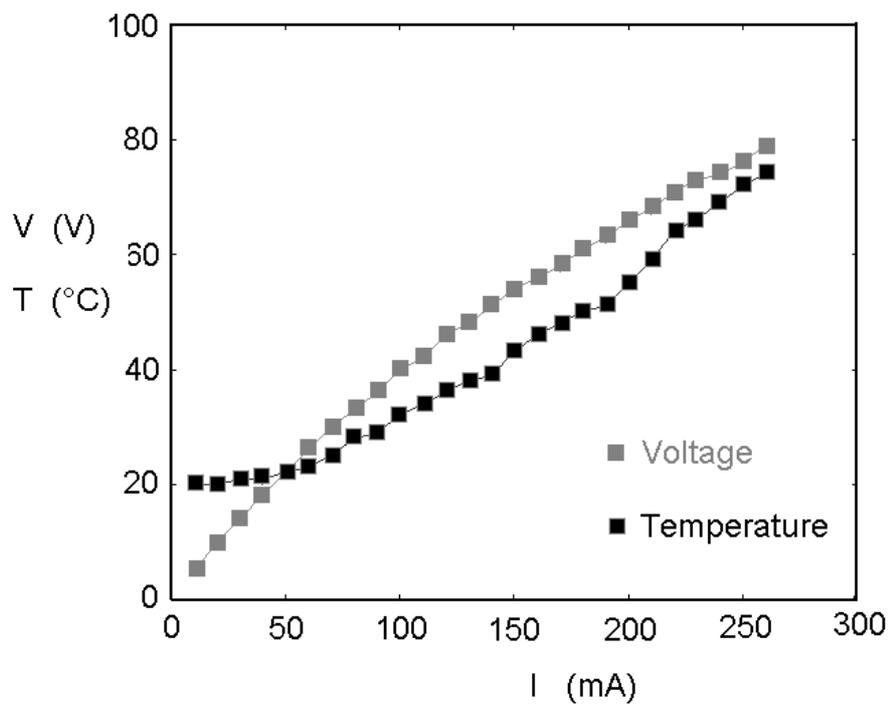

Fig.8
11

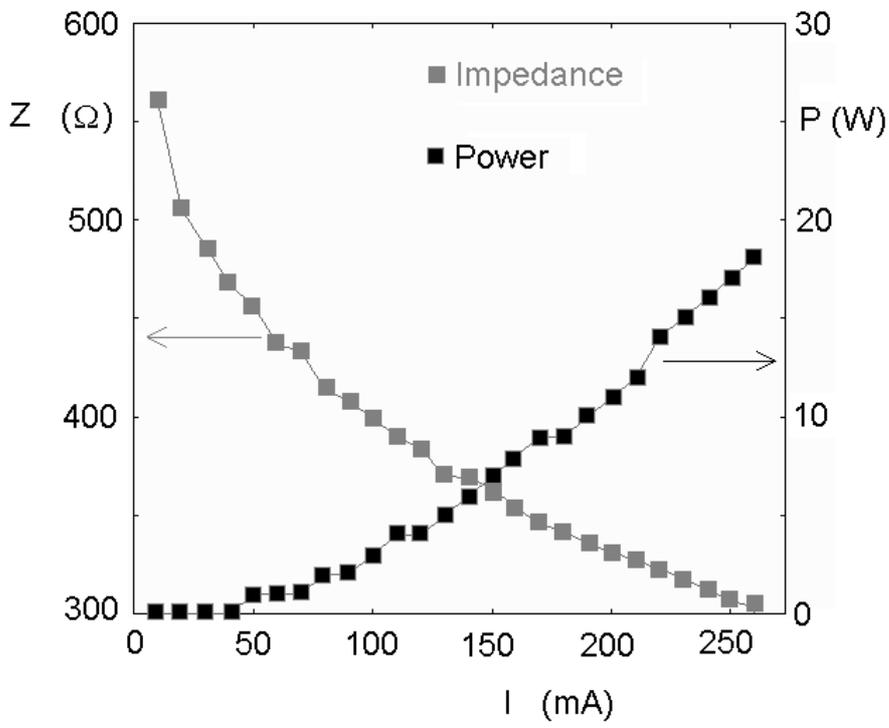

Fig.9